\documentclass[twocolumn, pra]{revtex4-1}

\usepackage{graphicx}
\usepackage{dcolumn} 
\usepackage{bm}
\usepackage{hyperref}
\usepackage{enumitem}
\usepackage[mathlines]{lineno}
\usepackage[titletoc, title]{appendix}

\graphicspath{ {Images/} }

\begin{document}

\title{Quantum dynamics of bosons in a two-ring ladder: dynamical algebra, vortex-like excitations and currents}

\author{Andrea Richaud}
\author{Vittorio Penna}
\affiliation{
Dipartimento di Scienza Applicata e Tecnologia and u.d.r. CNISM, Politecnico di Torino, 
Corso Duca degli Abruzzi 24, I-10129 Torino, Italy}
\date{\today}

\begin{abstract}
We study the quantum dynamics of the Bose-Hubbard model on a ladder formed by two rings coupled by tunneling effect.
By implementing the Bogoliubov approximation scheme, we prove that, despite the presence of the
inter-ring coupling term, the Hamiltonian decouples in many independent sub-Hamiltonians $\hat{H}_k$ associated to 
momentum-mode pairs $\pm k$. 
Each sub-Hamiltonian $\hat{H}_k$ is then shown to be part of a specific dynamical algebra. The properties
of the latter allow us to perform the diagonalization process, to find energy spectrum, the conserved quantities
of the model, and to derive the time evolution of important physical
observables. We then apply this solution scheme to the simplest possible closed ladder, the double
trimer. After observing that the excitations of the system are weakly-populated vortices, we explore the
corresponding dynamics by varying the initial conditions and the model parameters. Finally, we
show that the inter-ring tunneling determines a spectral collapse 
when approaching the border of the dynamical-stability region.   

\end{abstract}

\maketitle

\section{\label{sec:Introduction} Introduction}
Recent advances in ultracold atom Physics have made it possible to study a wide range of many-body quantum
systems of fermions, bosons and even mixtures of two atomic species. The phenomenology one can explore is
so rich that a new area of investigation, which goes under the name of Atomtronics, has emerged \cite{Olsen}.
A system of ultracold atoms subject to a spatially periodic potential features a band diagram which is conceptually
equivalent to those ones relevant to electrons in standard crystal lattices. It is therefore possible to engineer
ultracold atoms' equivalents of usual electronic materials, i.e. conductors, dielectrics and semiconductors
\cite{Seaman}. 
Properly tailoring the periodic optical potential, one can achieve behaviors similar to doped semiconductors. 
The latter can be used to realize the atomtronic counterpart of traditional electron devices, such as diodes and bipolar junction
transistors \cite{Transistors} which, in turn, can be used as building blocks for actual circuits, such as
amplifiers, flip-flops, and logic gates \cite{Pepino,Benseny}.  

In parallel, systems of ultracold neutral atoms have been used to simulate interesting many-body phenomena in
their most simple and essential forms, avoiding the complications usually encountered in actual materials 
\cite{Bloch,Feynman}. 
The charge neutrality of these systems does not prevent the observation of the 
interesting phenomena characterizing charged particles in a magnetic field. For example,
the equivalence between the Lorentz and the Coriolis force allows one to realize
synthetic magnetic fields in \textit{rotating} systems of neutral particles \cite{Lin}. 
Also, the current cutting-edge technologies have enabled the detection of bosonic chiral currents
in ladders \cite{Atala}, and the study of quantum transport in ultracold gases in optical lattices
\cite{Chien} and of topological quantum matter \cite{Goldman}.

In this work we focus on a specific lattice geometry, the Bose-Hubbard ladder with periodic boundary 
conditions. This kind of systems has been designed in \cite{Cataliotti} for a single ring and 
in \cite{Aghamalyan2013}
for a ladder, and has attracted increasing attention in the recent years. It consists of two vertically-stacked
rings whose sites are populated by weakly interacting bosons.
%
%
The current dynamics in a two-rings system subject to a synthetic magnetic field has been studied in \cite{Aghamalyan2013} in the weak-coupling regime by means of two-mode Gross-Pitaevskii equations.
The same mean-field approach has been used to study angular-momentum Josephson oscillations \cite{Wolf}
and the coherent transfer of vortices \cite{Gallemi}, while 
persistent currents flowing in the two-ring system have been demonstrated \cite{Amico_sci_rep} to 
provide a physical implementation of a qubit. The effect of an artificial magnetic field on an open ladder 
has been investigated in \cite{Orignac} and, more recently, in \cite{Piraud, Citro}. They have shown that 
this lattice geometry leads to the 1D- equivalent of a vortex lattice in a superconductor, and that a 
true Meissner to vortex transition occurs at a certain critical field. Finally, \cite{Tokuno} 
has presented a field-theoretical approach for the determination of the ground state, while 
different possible currents regimes have been studied in \cite{Uchino} and \cite{Haug}. 
Recently, the presence of the Meissner effect has been observed in the bosonic ladder 
\cite{Atala} while the phase diagram thereof has been discussed in \cite{Sachdeva}.

Motivated by the considerable interest in coupled annular Bose-Einstein condensates of recent years, 
in this paper we investigate the two-ring ladder from a different perspective, with the aim of giving 
an accurate insight into its quantum dynamics. We move from site-modes picture (where the expectations
values of operators are local order parameters of the lattice sites), to momentum-modes picture (where
expectation values of operators are collective order parameters in momentum space). In the momentum domain,
we perform the well-known Bogoliubov approximation under the assumption that in both rings the same momentum
mode $r$ is macroscopically occupied. The ensuing model Hamiltonian is shown to decouple into many
sub-Hamiltonians $\hat{H}_k$, one for each \textit{pair} of momentum modes. Each sub-Hamiltonian 
$\hat{H}_k$ is proved to belong to the dynamical Lie algebra so(2,3).  

The recognition of a certain dynamical algebra, together with its invariants, has been used to find
the spectrum and the time evolution of quantum systems \cite{Rasetti_generalized,ZFG,Kaushal,Rasetti_JJ,Penna_coupled_wells,Penna_central_depleted,Torrontegui,Penna_delocalization} and, once again, proves to be the key-element for the analytic solution of the model under
scrutiny. The remarkable importance of this abstract mathematical property is that it provides an
effective diagonalization scheme and helps to find conserved quantities. Moreover, the time evolution
of several meaningful observables belonging to the dynamical algebra
can be obtained by solving a linear system of differential equations.

In Section \ref{sec:Model_presentation}, we present the Bose-Hubbard (BH) Hamiltonian associated to
the two-ring ladder and implement the Bogoliubov scheme. In Section \ref{sec:Dynamical_algebra}, we
prove that the model Hamiltonian belongs to a dynamical algebra, namely so(2,3). We show that algebra
Casimir invariant correctly corresponds to angular momentum, and we find the excitation spectrum.
In Section \ref{sec:Double_trimer}, we apply this solution scheme to the simplest possible closed ladder,
the double trimer. Moreover, we show that the excitations of the system correspond to weakly populated vortices, 
we derive the time evolution of some physical observables commonly studied in the literature, and we describe 
various significant quantum processes that occur in the system. In Section \ref{sec:Vortex_dynamics}, 
we explore the dynamics of excited bosons by varying the initial conditions and the model parameters, 
emphasizing the role played by initial phase differences.
We also comment on the fact that properly choosing certain parameters, 
the system can approach dynamical instability. In particular, we show how a spectral collapse
takes place when the inter-ring tunneling reaches a specific critical value. 
Section \ref{sec:Concluding_remarks} is devoted to concluding remarks.

\section{\label{sec:Model_presentation}Model Presentation}
In this section we reformulate the BH Hamiltonian describing the ladder system
and including the inter-ring tunneling term by means of momentum modes
characterizing the Bogoliubov picture. 

\subsection{\label{sec:Site-modes_picture}Site-modes picture}
The second-quantized Hamiltonian describing bosons confined in a two-ring ladder is
$$
  \hat{H}=- T_a \sum_{j=1}^{M_s} \left(A_{j+1}^\dagger A_j +A_j^\dagger A_{j+1} \right) +\frac{U_a}{2} \sum_{j=1}^{M_s} N_j(N_j-1) 
$$
$$
    - T_b \sum_{j=1}^{M_s} \left(B_{j+1}^\dagger B_j +B_j^\dagger B_{j+1} \right) +\frac{U_b}{2} \sum_{j=1}^{M_s} M_j(M_j-1) 
$$
\begin{equation}
   -T \sum_{j=1}^{M_s} \left( A_j B_j^\dagger + B_j A_j^\dagger \right).
   \label{H1}
\end{equation}
One can recognize two \textit{intra}-ring tunnelling terms ($T_a$ and $T_b$), two on-site repulsive 
terms ($U_a$ and $U_b$), and an \textit{inter}-ring tunneling term $T$. These site-operators satisfy 
standard bosonic commutators: $[A_j,A^\dagger_k]=\delta_{j,k}$, $[B_j,B_k^\dagger]=\delta_{j,k}$ while 
$[A_j,B_k^\dagger]=0$. $N_j=A_j^\dagger A_j$ and $M_j=B_j^\dagger B_j$ are number operators. 
The number of lattice sites in each ring is denoted with $M_s$. 

\subsection{\label{sec:Momentum-modes_picture}
Momentum-modes picture} Due to the ring structure of the system, it is convenient to introduce momentum-mode 
operators $a_k$ and $b_k$, whose relation with sites operator is 
$$
 A_j=\sum_{k=1}^{M_s} \frac{a_k}{\sqrt{M_s}} e^{+i\tilde{k}aj}, \qquad B_j=\sum_{k=1}^{M_s} \frac{b_k}{\sqrt{M_s}} e^{+i\tilde{k}aj}, 
$$
with $\tilde{k}= \frac{2\pi}{L}k$ and $L = M_s a$. The length $a$ is the inter-site distance, $L$ is 
the ring circumference and the summations run on the first Brillouin zone. Notice that the use of the momentum-mode picture is justified by the fact that we are considering a \textit{repulsive} on-site interaction $U>0$, which, in turn, is linked to a ground state where bosons are delocalized in the system. 
Momentum-mode operators $a_k$ and $b_k$ inherit bosonic commutation relations: 
$[a_j,a_k^\dagger]=\delta_{j,k}$, $[b_j,b_k^\dagger]=\delta_{j,k}$ and $[a_j,b_k^\dagger]=0$. 
Number operators $n_k=a_k^\dagger a_k$ and $m_k=b_k^\dagger b_k$ count the number of bosons having
(angular) momentum $\hbar k$. In this new picture, the Hamiltonian can be written as
$$
  \hat{H}= \frac{U_a}{2M_s} \sum_{p,q,k=1}^{M_s}
  a_{q+k}^\dagger a_{p-k}^\dagger a_q a_{p} -2T_a\sum_{k=1}^{M_s} a_k^\dagger a_k \, \cos(a\tilde{k}) 
$$
$$
 +\frac{U_b}{2M_s} \sum_{p,q,k=1}^{M_s}    b_{q+k}^\dagger b_{p-k}^\dagger b_q b_{p} -2T_b\sum_{k=1}^{M_s} b_k^\dagger b_k \, \cos(a\tilde{k}) 
$$
$$
-T \sum_{k=1}^{M_s}\left( a_k b_k^\dagger+a_k^\dagger b_k\right). 
$$
Let us assume that, in both rings, momentum mode $r$ is macroscopically occupied. 
Further, for the sake of simplicity, we assume $M_s$ to be an odd (positive) integer. 
Under the hypothesis that the condensate is weakly interacting (small $U/T$), and
thus is in the superfluid region of the BH phase diagram,
it is possible to perform the well-known Bogoliubov approximation \cite{Rey}, 
\cite{Burnett} (see also Appendix \ref{sec:Bogoliubov_approximation} for details). 
We observe that this scheme can be applied as well in the case $U <0$, describing attractive bosons, 
provided the condition $|U|/T$ small enough, guaranteeing that bosons are delocalized and superfluid,
is fulfilled \cite{Jack}.
One thus discovers that the Hamiltonian, apart from a constant term, decouples in $(M_s-1)/2$ 
independent Hamiltonians $\hat{H}_k$, one for each \textit{pair} of momentum modes
\begin{equation}
\label{eq:decoupling}
\hat{H}= E_0 +\sum_{k> 0} \hat{H}_k
\end{equation}
where $E_0= {u_a}(N-1)/2+{u_b}(M-1)/2  -2\cos(a\tilde{r})(T_a  N + T_b M) -2T\sqrt{NM}$
is the ground-state energy, and
$$
\hat{H}_k= 2\sin(a\tilde{r})\sin(a\tilde{k}) \bigl[T_a \left( n_{r+k}-n_{r-k} \right )  
$$
$$
 + T_b \left( m_{r+k}-m_{r-k} \right ) \bigr]+
$$
$$
 + \gamma_{a,k} \left( n_{r+k} +n_{r-k}  \right) + u_a \left( a_{r+k}^\dagger a_{r-k}^\dagger  + a_{r+k} a_{r-k} \right)
$$
$$
+ \gamma_{b,k} \left( m_{r+k} +m_{r-k}  \right) + u_b \left( b_{r+k}^\dagger b_{r-k}^\dagger  + b_{r+k} b_{r-k} \right)
$$
$$
-T\left( a_{r+k}b_{r+k}^\dagger +a_{r+k}^\dagger b_{r+k} + a_{r-k}b_{r-k}^\dagger +a_{r-k}^\dagger b_{r-k} \right).
$$
Parameters
$$
 \gamma_{a,k} = -2T_a \cos(a\tilde{r})\bigl(\cos(a\tilde{k})-1\bigr) +u_a -T\sqrt{\frac{M}{N}},
$$
$$
 \gamma_{b,k} = -2T_b \cos(a\tilde{r})\bigl(\cos(a\tilde{k})-1\bigr) +u_b -T\sqrt{\frac{N}{M}},
$$
$$ 
 u_a = \frac{U_a N}{M_s},\qquad u_b = \frac{U_b M}{M_s},
$$
have been introduced to simplify the notation, and $N$ and $M$ are the total number of bosons in the two rings.
If $T_a=T_b=T_\parallel$, the whole term 
\begin{equation}
2\sin(a\tilde{r})\sin(a\tilde{k}) T_\parallel  \left( n_{r+k}-n_{r-k} + m_{r+k}-m_{r-k} \right )
\label{eq:Constant_of_motion}
\end{equation}
can be shown (see Section \ref{sec:Dynamical_algebra}) to be a constant of motion and thus can be incorporated in $E_0$. The natural basis of the Hilbert space relevant to model Hamiltonian $\hat{H}_k$ is $\left\{| n_{r+k},\,n_{r-k},\,m_{r+k},\,m_{r-k}\rangle \right\}$, a basis vector being labelled by \textit{four} momentum quantum numbers. As regards values $n_0$ and $m_0$, due to Bogoliubov approach, they inherently depend on $n_{r\pm k}$ and $m_{r\pm k}$, their expression being $n_0=N-\sum_k \left( n_{r+k}+n_{r-k}\right)$ and $m_0=M-\sum_k \left( m_{r+k}+m_{r-k}\right) $.

\section{Dynamical Algebra}
\label{sec:Dynamical_algebra}
In general, a dynamical algebra $\mathcal{A}$ is a Lie algebra, i.e. $n-$dimensional vector space spanned by $n$ generators (operators) $\hat{e}_1, \hat{e}_2, \dots, \hat{e}_n$ closed under commutation. The closure property means that the commutator of any two algebra elements is again an algebra element. A Lie algebra is univocally specified once all the commutators $\left[\hat{e}_j, \hat{e}_k\right]=i\sum_m f_{jkm}\, \hat{e}_m$ are given, namely when the set of the so called structure constants $\left\{f_{jkm}\right\}$ is specified \cite{ZFG}.
A model Hamiltonian $\hat{H}$ belongs to a dynamical algebra $\mathcal{A}=\mathrm{span}\{\hat{e}_1, \hat{e}_2, \dots, \hat{e}_n\} $ whenever $\hat{H}$ can be expressed as a linear combination $\hat{H}=\sum_j h_j \hat{e}_j$ of the generators of $\mathcal{A}$.  The important consequences of this property are that 
\begin{enumerate}[label=(\alph*)]
\item Conserved physical quantities correspond to algebra's invariants, 
\item The diagonalization process of $\hat{H}_k$ becomes straightforward,
\item The Heisenberg equations
can be shown to form a simple linear system of differential equations.
\end{enumerate}
Under the assumption $T_a = T_b = T_\parallel$, Hamiltonian $\hat{H}_k$ is recognized to be an element
of the dynamical algebra $\mathcal{A}=$ so(2,3), a $10-$dimensional Lie algebra spanned by operators 
$$
A_+ = a_{r+k}^\dagger a_{r-k}^\dagger,
\quad
B_+ = b_{r+k}^\dagger b_{r-k}^\dagger,
$$ 
$$
A_- = (A_+)^\dagger, \quad B_- =(B_+)^\dagger,
$$
\begin{equation}
\label{eq:A3B3}
 A_3 = \frac{n_{r+k}+n_{r-k}+1}{2}, \quad B_3 = \frac{m_{r+k}+m_{r-k}+1}{2},
\end{equation}
$$
 S_+ = a_{r+k}^\dagger b_{r+k}+a_{r-k}^\dagger b_{r-k}, \quad S_- = (S_+)^\dagger,
$$
$$
 K_+ = a_{r-k}^\dagger b_{r+k}^\dagger +a_{r+k}^\dagger b_{r-k}^\dagger, \quad K_- = (K_+)^\dagger \; .
$$
One can easily see that $\hat{H}_k$, up to an inessential constant quantity 
$-\sum_{k>0}(\gamma_{a,k}+\gamma_{b,k})$,
can be written as
$$
 \hat{H}_k = 2\gamma_{a,k} A_3 +u_a \left( A_+ +A_-\right)+2\gamma_{b,k} B_3 +u_b \left( B_+ +B_-\right)
$$
$$
-T\left(S_++S_-\right)
$$
where operators $\left\{A_+,\,A_-,\,A_3 \right\}$, associated to the ring A, generate a 
su(1,1) algebra marked by 
the well-known commutators $\left[A_+,A_-\right]=-2A_3$, $\left[A_3, A_\pm \right ]= \pm A_\pm$, and 
operators $\left\{B_+,\,B_-,\,B_3 \right\}$, relevant to ring B, feature the same su(1,1) structure (an application of this dynamical algebra can be found in \cite{Solomon} for a trapped condensate). 

However, the important term in ${\hat H}_k$ is the inter-ring tunnelling term which is responsible for
an algebraic structure considerably more complex than the simple direct sum of two su(1,1) algebras. 
In Appendix \ref{sec:Defining_commutators}, the commutators of $A_\pm$, $B_\pm$, $A_3$, 
$B_3$, $K_\pm$, and $S_\pm$ are explicitly calculated showing that indeed they form an algebra so(2,3). 

\subsection{The algebra invariant as a constant of motion}
\label{sub:Algebra_invariant}
 In the absence of the inter-ring tunnelling term, i.e. if $T$ is zero, the two rings decouple and $\hat{H}_k$ could be seen as an element of the direct sum of two commuting algebras su(1,1). In such a case, the difference between the number of bosons having momentum $r+k$ and the number of bosons having momentum $r-k$ is a conserved quantity in each single ring. This statement can be easily proved by using the Casimir operator of algebra su(1,1) for ring A 
 $$
    C_a = A_3^2-\frac{1}{2}\left(A_+A_-+A_-A_+\right)= A_4(A_4+1),
 $$
where 
$$
    A_4=\frac{n_{r+k}-n_{r-k}-1}{2}.
$$
We recall that, by definition, the Casimir operator (or, equivalently $A_4$) commute with all the algebra generators
$A_\pm$, $A_3$. The same comment holds for the Casimir operator $C_b$ of $B_\pm$ and $B_3$.

Conversely, in the presence of the inter-ring tunnelling term, neither $n_{r+k}-n_{r-k}$ 
nor $m_{r+k}-m_{r-k}$ any longer represent conserved quantities. Nevertheless, by applying
the general recipe described in \cite{Gilmore}, one discovers that 
the Casimir operator of $\mathcal{A}$= so(2,3) is 
$$
C = C_a +C_b + \frac{S_+S_- +S_-S_+}{4}-\frac{K_+K_- +K_-K_+}{4} .
$$
This operator, a quadratic form involving all the algebra elements, can be rewritten in
the standard form
$$
  C= C_4 \left(C_4+2 \right)
$$
where
$$
    C_4 =\frac{n_{r+k}-n_{r-k}+m_{r+k}-m_{r-k}}{2} -1\; .
$$
The conserved quantity $L_z (k)= n_{r+k}-n_{r-k}+m_{r+k}-m_{r-k}$ has a nice physical interpretation. 
Apart from the inessential additive constant $-1$, $C_4$ is proportional to the difference 
between the numbers of bosons having momentum $r+k$ and momentum $r-k$ in the whole ring ladder.
Then $L_z(k)$ can be interpreted as the angular momentum for the modes $r\pm k$.
This fact not only proves the ansatz on the constant of motion (\ref{eq:Constant_of_motion}) but, since 
it holds for every sub-Hamiltonian $\hat{H}_k$, leads to the natural conclusion that the 
angular momentum $L_z= \sum_{k>0} L_z(k)$ of the whole system is a conserved quantity. 
In this regard, it is worth noting that the ten operators which generate algebra so(2,3) always correspond to two-bosons processes where angular momentum is conserved.

\subsection{Spectrum and diagonalization}
\label{sub:Spectrum}
Once the dynamical algebra has been identified, the Hamiltonian
$\hat{H}_k$ can be diagonalized thanks to a simple unitary transformation 
$U$ of group SO(2,3) defined as
$$
 U=  e^{\frac{\varphi}{2}(S_- -S_+)}
    e^{\frac{\xi}{2}(K_--K_+)}
    e^{\frac{\theta_a}{2}(A_- -A_+)} e^{\frac{\theta_b}{2}(B_- -B_+)}.
$$
This represents the central step of the dynamical-algebra method.
A suitable choice of parameters $\varphi$, $\xi$, $\theta_a$ and $\theta_b$, 
(see Appendix \ref{sec:Rotation_angles} for their explicit expressions), 
allows to write the Hamiltonian 
$$
    \hat{\mathcal{H}}_k = U^{-1}\,\hat{H}_k\,U=  \bigg[c_1\cosh \theta_a  -c_2 \sinh \theta_a \bigg]A_3 +
$$
$$
+ \bigg[c_3 \cosh\theta_b -c_4\sinh\theta_b  \bigg] B_3
$$
as a linear combination of $A_3$ and $B_3$, operators which are \textit{diagonal} in the Fock-states basis.
The explicit expression of coefficients $c_1$, $c_2$, $c_3$ and $c_4$ is given in Appendix \ref{sec:Rotation_angles}
showing that they are complex functions of the interaction and the tunneling parameters. Based on the definitions (\ref{eq:A3B3}), the spectrum of Hamiltonian $\hat{H}_k$ is found to be
$$
E_k (n_{r+k} ,n_{r-k}, m_{r+k} ,m_{r-k}) =  
$$
$$
+ \bigg[c_1\cosh \theta_a  -c_2 \sinh \theta_a \bigg]
\frac{n_{r+k} +n_{r-k} +1}{2}
$$
$$
+ \bigg[c_3 \cosh\theta_b -c_4\sinh\theta_b  \bigg] \frac{m_{r+k} +m_{r-k} +1}{2}
$$
where $n_{r\pm k}$ and  $m_{r\pm k}$ now represent the quantum numbers describing the boson populations.

\subsection{The time-evolution of algebra elements}
\label{sub:Linear_system}
The knowledge of the dynamical algebra $\mathcal{A}$ relevant to a given model Hamiltonian $\hat{H} = \sum_j h_j \hat{e}_j$ allows one to derive in a direct way the equations of motion of any physical observable $\mathcal{O} = \sum_k o_k \hat{e}_k$ written in terms of the generators $\hat{e}_k \in \mathcal{A}$. If $[\hat{e}_j, \hat{e}_k]= i \sum_m f_{jkm} \hat{e}_m$ represent the commutators 
of $\mathcal{A}$ ($f_{jkm}$ are the algebra structure constants),
then the Heisenberg equation for $\hat{e}_k$ reduces to a simple linear combination of the generators 
$$
i \hbar \frac{\mathrm{d}}{\mathrm{d}t} \hat{e}_k = [ \hat{e}_k , \hat{H} ] 
= i \sum_m \rho_{km} \hat{e}_m , 
$$
where $\rho_{km} =\sum_j  h_j  f_{jkm}$ and the commutators have been used to explicitly calculate $[ \hat{e}_k , {\hat H} ]$.
The dynamical evolution of the whole system is thus encoded in a simple set of linear equations whose 
closed form is ensured by the commutators of the dynamical algebra, and whose number corresponds to the algebra dimension. 
The evolution of physical observables $\mathcal{O} $ are thus fully determined by the one of generators $\hat{e}_k$.

Concerning the dynamical algebra so(2,3) the linear system of differential equations is: 
$$
    i\hbar \dot{A}_3 =  u_a(A_+-A_-)-T\left(\frac{1}{2}S_+ - \frac{1}{2}S_-\right ),
$$
$$
    i\hbar\dot{A}_-= 2\gamma_{a,k}A_- +2u_a A_3-TK_-,
$$
$$
    i\hbar \dot{B}_3 = u_b(B_+-B_-)-T\left(-\frac{1}{2}S_+ + \frac{1}{2}S_-\right ),
$$
$$
    i\hbar\dot{B}_-= 2\gamma_{b,k}B_- +2u_b B_3-TK_-,
$$
$$
    i\hbar\dot{S}_-= \gamma_{a,k}  S_- +u_aK_+ -  \gamma_{b,k} S_- -u_b K_- +2T(A_3-B_3),
$$
$$
    i\hbar\dot{K}_- =  \gamma_{a,k} K_- +u_a S_+ + \gamma_{b,k} K_- +u_b S_- -2T(A_- + B_-).
$$
Of course the remaining four equations for $A_+$, $B_+$, $S_+$ and $K_+$ are the hermitian conjugates of the Heisenberg equations for $A_-$, $B_-$, $S_-$ and $K_-$. Rigorously, this is a system of \textit{operator} ordinary differential equations (ODEs), as the unknowns are the time evolution of operators. In the following we will switch from operators to their expectation values, i.e. from operator ODEs to standard complex ODEs. In fact the structure of Heisenberg equations remains unchanged when taking the expectation values on both sides, e.g.,

$$
 i\hbar \frac{\mathrm{d}}{\mathrm{d}t}\langle A_3\rangle  = u_a \left(\langle A_+  \rangle  - \langle A_-  \rangle \right) -T\left( \frac{1}{2} \langle S_+  \rangle - \frac{1}{2}\langle S_-  \rangle\right).
$$
This conceptual jump will be often made and always understood in all this manuscript.

\section{Double trimer}
\label{sec:Double_trimer}
The formulae we have presented so far are very general because they can capture the dynamics 
of physical regimes distinguished by an arbitrary macroscopic mode $r$ with $0\le r \le M_s-1$
and an arbitrary choice of the site number $M_s$ and of the other model parameters. In particular, 
for $r \ne 0$, our approach allows to investigate the dynamics of quantum excitations relevant
to the  macroscopic (semiclassical) double-vortex state 
characterized by a total vorticity proportional to $r$
$$
A_j (t, r) = \sqrt{\frac{N}{M_s}} e^{ ij{\tilde r} -\omega_r t}, 
\qquad B_j (t, r) = \sqrt{\frac{M}{M_s}} e^{ ij{\tilde r} -\omega_r t},
$$
where $A_j$, $B_j$ are the local order parameters (of the semiclassical Hamiltonian) associated
to model (\ref{H1}), and $\omega_r$ can be found by the corresponding dynamical equations 
\cite{CP2002}.
In this section we show a simple and yet very interesting application of
the solution scheme we have proposed. We consider the smallest possible ladder, the one formed by two rings with $M_s=3$ sites (trimer). This systems has received a considerable attention
in the last decade in that it represents the minimal circuit in which chaos can be triggered 
\cite{BFP2009,Arwas2015,Han2016}. 
Moreover, we assume that the macroscopically occupied
mode is $r=0$ (entailing that no macroscopic current is present), that upper and lower ring host
an equal number of bosons ($N=M$) and that the intra-ring tunnelling and the on-site repulsion
parameters are equal ($T_a=T_b=:T_\parallel$ and $U_a=U_b=:U$). As a consequence,
$\gamma_{a,k}=\gamma_{b,k}=\gamma_k$. Since the first Brillouin zone involves just 
three modes $k=-1,\,0,\,1$, then there is only one $\gamma_k$ that will be denoted by
$$
 \gamma=\gamma_1= 2T_\parallel \left[1-\cos\left(\frac{2\pi}{3 }\right)\right] +u-T .
$$
By performing the Bogoliubov approximation with the momentum mode $r=0$ macroscopically occupied, 
the double-trimer Hamiltonian can be written as $\hat{H}=E_0+\hat{H}_1$, where
$$
E_0= u(N-1)-2N(2T_\parallel +T) -2\gamma ,
$$
$$
    \hat{H}_1 = 2 \gamma  (A_3+B_3) +u(A_++A_-+B_++B_-) - T(S_++S_-) .
$$
Notice that $E_0$ is a constant quantity, while $\hat{H}_1$ 
rules the dynamics of bosons having wave-number $k=\pm 1$. 

As we have proved in the previous section, the total angular momentum, which is proportional 
to $n_1-n_{-1}+m_1-m_{-1}$, is a conserved quantity. In passing, we note that for a double trimer with twin
rings ($\gamma_a = \gamma_b$, $u_a=u_b$), a second quantity commuting with Hamiltonian $\hat{H}_1$ 
$$
 I =\left( a_1 b_1^\dagger +a_1^\dagger b_1\right) 
 -\left(a_{-1}b_{-1}^\dagger + a_{-1}^\dagger b_{-1} \right),
$$
can be found.
In view of the mode coupling characterizing the hopping term of a BH model, $I$ can be interpreted as the difference between the tunnelling energies associated to the ring-ring boson exchange for the bosons of mode $k=1$,
and the bosons of mode $k=-1$.

Concerning the diagonalization of $\hat{H}_1$, generalized rotation angles $\varphi$, $\xi$, $\theta_a$ and $\theta_b$ are, in such a system
\begin{equation}
\label{eq:Rotation_angles}
\varphi=\frac{\pi}{2}, \quad \xi=0, 
\quad {\rm th}\; \theta_{a}= \frac{u}{\gamma - T},
\quad {\rm th}\; \theta_{b}= \frac{u}{\gamma + T}.
\end{equation}
They lead to the diagonal Hamiltonian
\begin{equation}
\label{eq:Diagonal_Hamiltonian}
   \hat{\mathcal{H}}=    E_0 +2 \hbar \omega \; A_3 + 2 \hbar \Omega \;B_3
\end{equation}
where the two frequencies 
\begin{equation}
    \label{eq:Char_freq_1}
     \omega = \frac{\sqrt{3T_\parallel(3T_\parallel+2u)}}{\hbar},
\end{equation}
\begin{equation}
      \label{eq:Char_freq_2}
      \Omega =\frac{ \sqrt{(3T_\parallel-2T)(3T_\parallel-2T+2u)}}{\hbar} ,
\end{equation}
have been defined.
Since $A_3 = (n_{k} +n_{-k}+1)/2$, and $B_3 = (m_{k} +m_{-k}+1)/2$, then $\hat{\mathcal{H}}$
formally corresponds to a system of four independent harmonic oscillators
with the spectrum $E_k (n_{k}, n_{-k}, m_{k}, m_{-k}) =E_0 + \hbar \omega (n_{k} +n_{-k}+1) 
+ \hbar \Omega (m_{k} +m_{-k}+1)$. 
%
The angle $\theta_a$ and the argument of the square roots are well defined only in a certain region of the three dimensional parameter space $(T_\parallel,\,U,\,T)$. From a dynamical point of view, approaching the border of this
{\it stability region} implies that the system tends to be unstable and many physical quantities manifest
diverging behaviors. This issue will be addressed in Section \ref{sec:Vortex_dynamics}.

\subsection{Vortex-like excitations and currents}
It is interesting to notice that the weak excitations in each ring are weakly-populated vortices. To show it, let us observe that, in the semi-classical picture, site-mode operators corresponding to momentum-mode operators $a_0=\sqrt{N-n_1-n_{-1}}$, $a_1=\sqrt{n_1}e^{i\phi_1}$, $a_{-1}=\sqrt{n_{-1}}e^{i\phi_{-1}}$ are
$$
A_j= \frac{1}{\sqrt{3}} \left[\sqrt{n_0} 
 + e^{+i\frac{2\pi}{3}j} \sqrt{n_{-1}}e^{i\phi_{-1}} + e^{-i\frac{2\pi}{3}j}   \sqrt{n_{1}}e^{i\phi_{1}} \right]
$$
where $n_0=N-n_{-1}-n_1$ and $j=1, 2, 3$ is the site index. 
The structure of site operators $A_j$'s, whose expectation values are the local order parameters, clearly shows that the state of the system is the superposition of three contributions, namely, a major mode $a_0$ corresponding to a zero super-current, and two minor modes ($a_1$ and $a_{-1})$ corresponding to counter-rotating weakly populated vortices. The same holds also for site-mode operators $B_j$ of ring B.

\begin{figure}[h!]
\includegraphics[width=1\columnwidth]{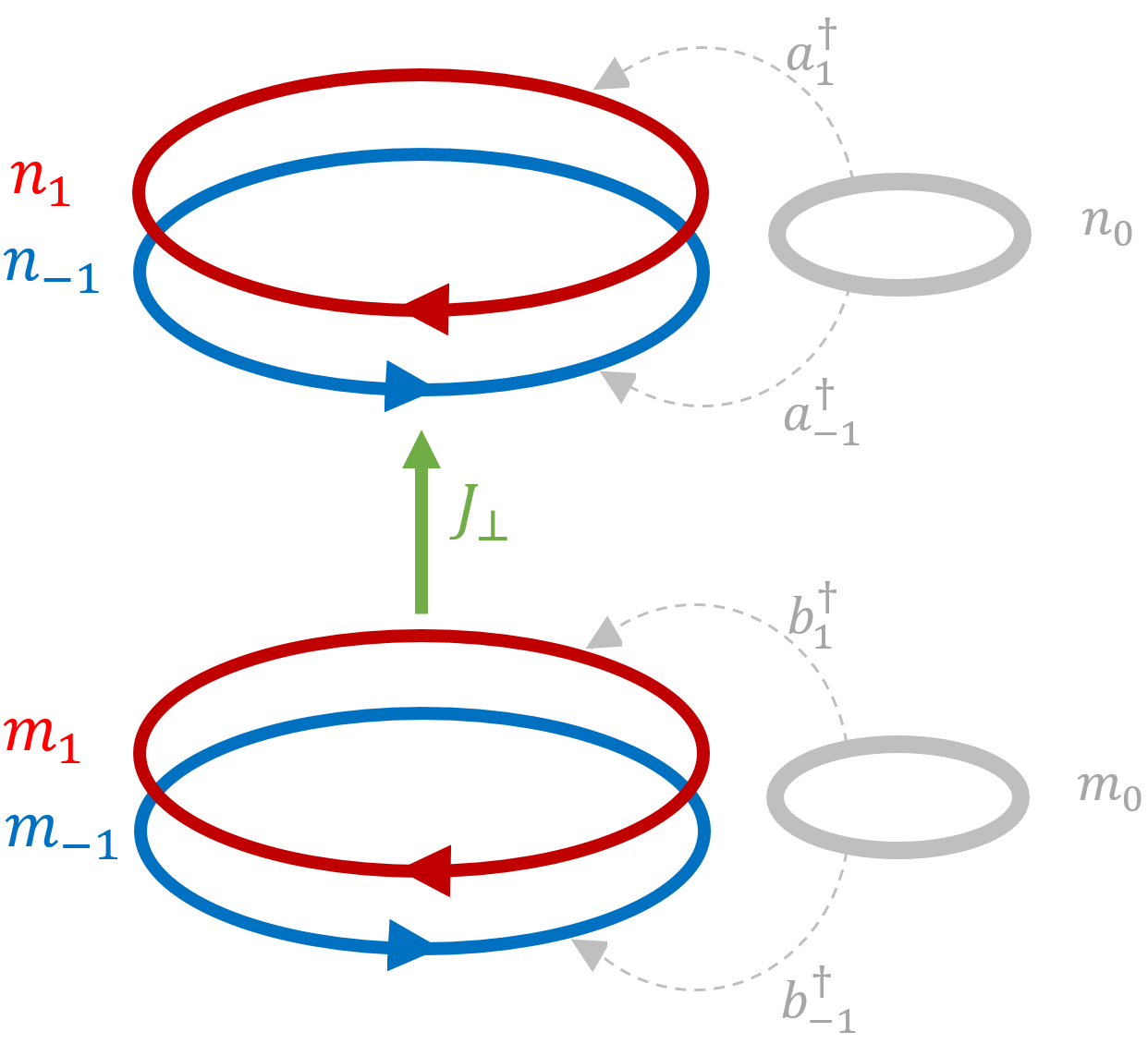}
\centering
\caption{Schematic representation of the vortex-like weak excitations in our system. In each ring there can be both a clockwise current and an anti-clockwise current. $J_\perp$ denotes the current of excited bosons between the two rings. The macroscopically occupied modes $n_0$ and $m_0$ (in gray), being semi-classic, can be considered as a sort of reservoir. }
\label{fig:Illustrazione_correnti}
\end{figure}

Let us introduce some observables, commonly found in literature (see for example \cite{Natu,Piraud})
whose time evolution allows one to illustrate the significant transport phenomena and inter-ring exchange
processes occurring in the system. We start with the currents along the two rings
$$
 J_A = i T_\parallel \sum_{l=1}^{3} \bigl(A^\dagger_{l+1}A_l-A^\dagger_l A_{l+1} \bigr)=\sqrt{3}T_\parallel \left(n_1 -n_{-1} \right),
$$
$$
 J_B = i T_\parallel \sum_{l=1}^{3} \bigl(B^\dagger_{l+1}B_l-B^\dagger_l B_{l+1} \bigr)=\sqrt{3}T_\parallel \left(m_1 -m_{-1} \right).
$$
These currents are proportional to the angular momenta in each single ring. Their superposition $J_{tot}=J_A+J_B$ 
is proportional to the total angular momentum (and thus is a conserved quantity), while $J_{chir}=J_A-J_B$ is proportional to the angular-momentum difference between the two rings. $N_*=n_1+n_{-1}+m_{1}+m_{-1}$ is the total number of excited bosons. The rung excitations' current
$$
  J_{\perp} = i T \sum_{l=1}^{3} \bigl(A^\dagger_{l}B_l-B^\dagger_l A_{l} \bigr)
$$
$$
= iT\bigl(a_1^\dagger b_1 + a_{-1}^\dagger b_{-1} - a_1 b_1^\dagger -a_{-1} b_{-1}^\dagger \bigr)
$$
measures the flow of excited bosons from ring B to ring A. Figure \ref{fig:Illustrazione_correnti} sketches the scenario of physical observables which come into play.

\subsection{Time evolution of observables and dynamical algebra}
Based on the scheme described in Section \ref{sub:Linear_system}, the dynamical algebra so(2,3) allows one to determine
the equations of motion for the total number of excited bosons $N_*$ and for the rung current $J_\perp$.
Since these two observables can be written as linear combinations of the so(2,3) generators, it is possible to write their dynamical equations in terms of algebra elements. Concerning $N_*$, the corresponding Heisenberg equation
is found to be
$$
\frac{\mathrm{d} N_*}{\mathrm{d}t} =2\left(\dot{A}_3+\dot{B}_3 \right)=
-\frac{2 u}{\hbar}i\left(A_+-A_-+B_+-B_-\right).
$$
Recalling that $A_+-A_-+B_+-B_-=a_1^\dagger a_{-1}^\dagger - a_1 a_{-1 } + b_1^\dagger b_{-1}^\dagger - b_1 b_{-1 }$,
this equation shows that the time variation of $N_*$ is proportional to a generalized current which can be interpreted as the boson-pairs flow from the macroscopically occupied modes $a_0$ and $b_0$ to the excited modes. In other words, this is the generation rate of boson pairs populating modes $k=\pm 1$. 
Such a pair is created by $a_1^\dagger a_{-1}^\dagger$ (annihilated by $a_1a_{-1}$) extracting (releasing) bosons
from (to) the macroscopically occupied modes which, being semi-classical, have bowed out and act as reservoirs (see Figure \ref{fig:Illustrazione_correnti}).  

As regards the rung current, the relevant Heisenberg equation reads
$$
   \frac{\mathrm{d}}{\mathrm{d}t} j_\perp = iT \left( \dot{S}_+-\dot{S}_- \right) =  \frac{-4T^2}{\hbar} \left(A_3-B_3\right) = 
$$
$$
 =\frac{-2T^2}{\hbar} \left(n_1 +n_{-1} - m_1-m_{-1} \right) 
$$
showing that a populations imbalance between the two rings is responsible for the time variation of the rung current.  

If one is interested in obtaining finer-grained info about the system, e.g., in finding the time evolution of a certain population $n_{\pm 1}$ or $m_{\pm 1}$, one must consider an enlarged dynamical algebra containing the original framework so(2,3). This is represented by the 15-dimensional algebra so(2,4) which, in fact, includes the 10-dimensional algebra so(2,3). It is within this enlarged algebra (whose generators are listed in Appendix \ref{sec:Commutators_so(2,4)}) that the dynamics of all the previously presented observables can be represented. The time evolution of excited populations
are easily found to be
$$
    i\hbar \dot{n}_1 = u (a^\dagger_1 a^\dagger_{-1} -a_1 a_{-1}) - T( - a_1b_1^\dagger +a^\dagger_1b_1),
$$
$$
    i\hbar \dot{n}_{-1} = u (a^\dagger_1 a^\dagger_{-1} -a_1 a_{-1}) - T(-a_{-1}b_{-1}^\dagger +a_{-1}^\dagger b_{-1} ),
$$
$$
    i\hbar \dot{m}_1 = u (b^\dagger_1 b^\dagger_{-1} -b_1 b_{-1})-T(a_1 b_1 ^\dagger -a^\dagger_1 b_1),
$$
$$
    i\hbar \dot{m}_{-1} = u (b^\dagger_1 b^\dagger_{-1} -b_1 b_{-1})-T(a_{-1} b_{-1} ^\dagger -a^\dagger_{-1} b_{-1}).
$$
These equations clearly show that the time evolution of excited populations can be triggered either by intra-ring processes $(u)$ or by inter-ring tunnelling $(T)$. Eventually, the time evolution of the chiral current is easily 
found to be
$$
  \frac{\mathrm{d}}{\mathrm{d}t}\, J_{chir} = i \frac{2\sqrt{3}}{\hbar} T_\parallel T  \left(a_1^\dagger b_1 +a_{-1}b_{-1}^\dagger  -a_1 b_1^\dagger -a_{-1}^\dagger b_{-1}     \right) .
$$
This equation confirms the intuitive fact that the angular-momentum difference between the two rings cannot evolve in time if the inter-ring tunnelling parameter $T$ tends to zero. 

As already noticed, the diagonal Hamiltonian (\ref{eq:Diagonal_Hamiltonian}) and the dynamics of physical
observables we have presented are featured by two characteristic frequencies $\omega$ and $\Omega$
which, in the limit $T\rightarrow 0$, turn out to be equal.
Different choices of parameters $\left(T_\parallel,\,u,\,T\right)$ result in different physical regimes, an aspect that will be discussed in the next section.
\bigskip

%
\onecolumngrid

\begin{figure}[h!]
\includegraphics[width=1\columnwidth]{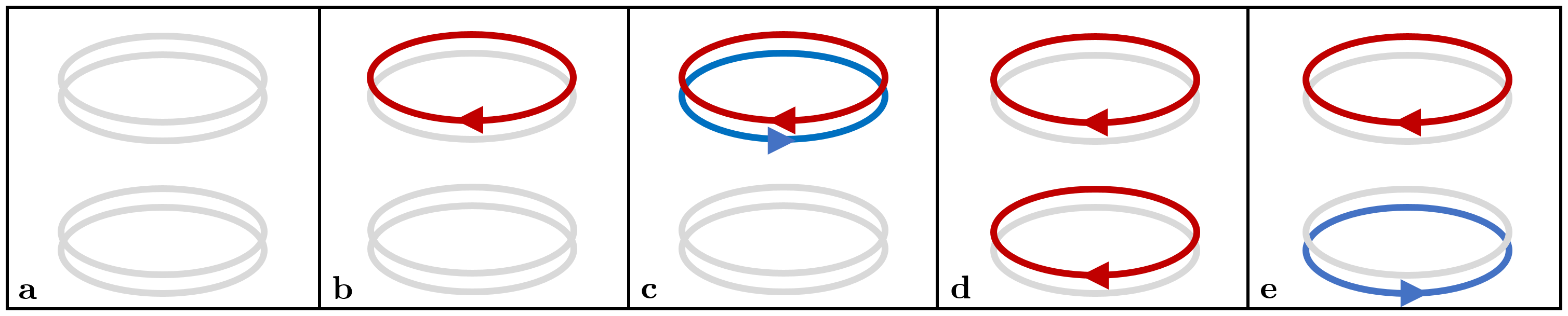}
\centering
\caption{Five different initial conditions. a) No initial excitations, b) A weak vortex in one ring, c) A pair of counter rotating weak vortices in the same ring, d) Two equal weak vortices in the two rings, e) A pair of counter rotating weak vortices in the two rings.}
\label{fig:Sketch_riepilogativo}
\end{figure}

\twocolumngrid

\section{Vortexlike-excitations and currents dynamics} 
\label{sec:Vortex_dynamics}
With reference to the double trimer, in the semi-classical picture, the expectation values of momentum-mode operators (expressed in terms of complex order parameters) can be written as
$$
   a_1=\sqrt{n_1}e^{i\phi_1}, 
   \qquad
   a_{-1}=\sqrt{n_{-1}}e^{i\phi_{-1}},
$$
$$
    b_1=\sqrt{m_1}e^{i\psi_1}, 
    \qquad 
    b_{-1}=\sqrt{m_{-1}}e^{i\psi_{-1}}.
$$
In this section we show how different initial conditions (i.e. \textit{moduli} and \textit{phases} of the aforementioned operators at $t=0$) together with different choices of parameters $T_\parallel$, $u$, and $T$ lead to different dynamical regimes. The explicit solutions of Heisenberg equations giving the time evolution of excited populations ($n_{\pm 1}(t)$ and $m_{\pm 1}(t)$) and of the rung current $J_\perp(t)$ can be found in the Supplemental Material. 
Figure \ref{fig:Sketch_riepilogativo} sketches the five different initial conditions we will focus on.

\paragraph{No initial excitations.} If, at $t=0$, $n_1=n_{-1}=m_1=m_{-1}=0$ meaning that no excitations are present in the initial state, as time goes on, excitations pairs are periodically created and annihilated according to the relations 
$$
     n_{\pm 1}(t)=m_{\pm 1}(t)= \frac{1}{2}\frac{u^2}{\hbar^2} \left[\frac{\sin^2\left(\omega\, t\right)}{\omega^2} + \frac{\sin^2\left(\Omega\, t\right)}{\Omega^2}\right].
$$
This example clearly shows how a non-zero on-site repulsive term $u$ determines fluctuations of the vacuum state $|n_1,n_{-1},m_1,m_{-1}\rangle=|0,0,0,0\rangle$. Chiral and rung currents are identically zero.
\begin{figure}[h!]
\includegraphics[width=1\columnwidth]{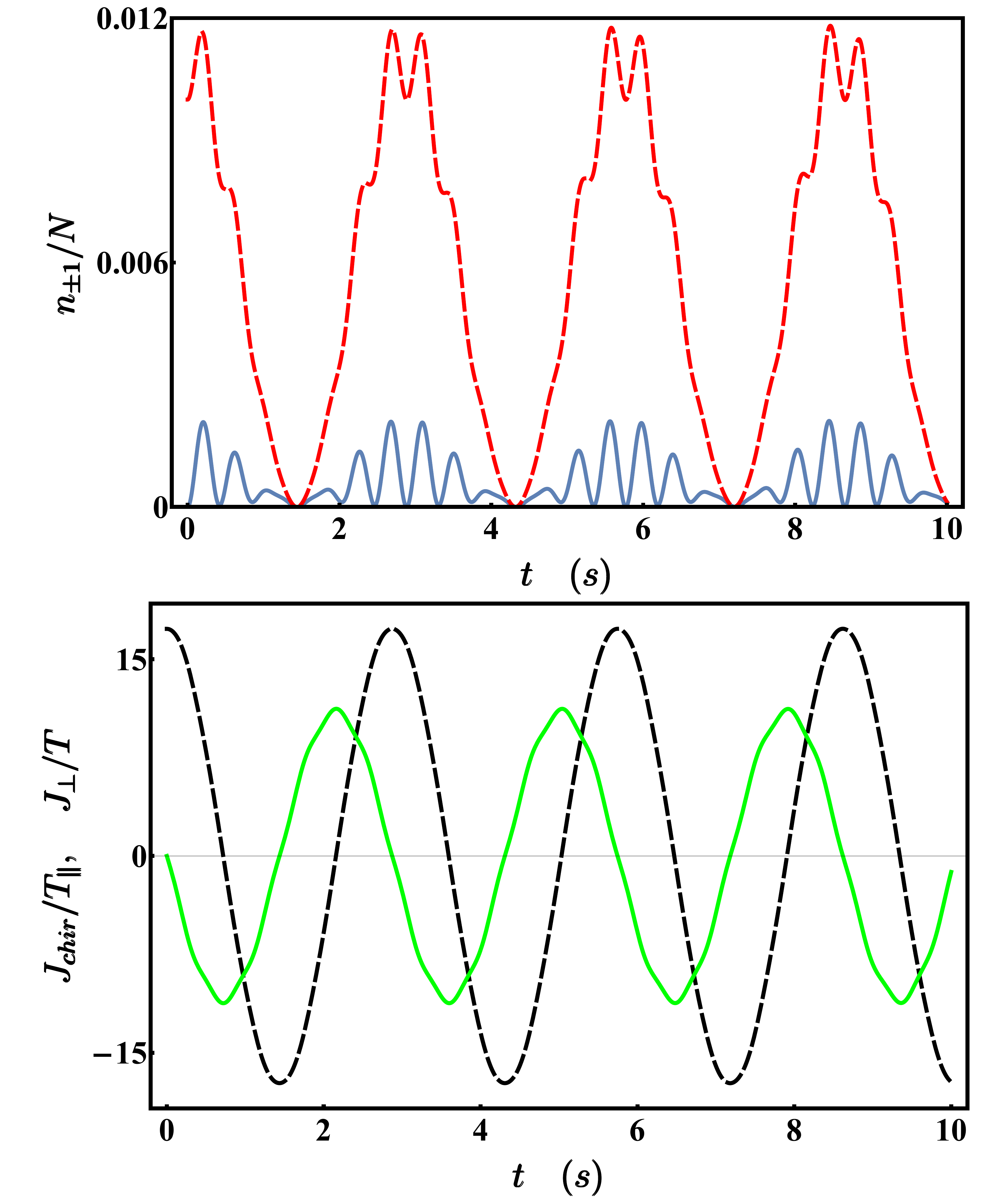}
\centering
\caption{Populations and currents dynamics for $T_\parallel=2$, $U=0.01$, $T=1$, $\hbar=1$, $N=1000$, $n_1(0)=10$ and $n_{-1}(0)=m_{\pm 1}(0)=0$. Upper panel: $n_{1}(t)$ corresponds to red dashed line while $n_{-1}(t)$ corresponds to blue solid line. $m_{1}(t)$ and $m_{-1}(t)$ feature the same behaviour but are shifted of a semi-period. Lower panel: $J_{chir}(t)$ is depicted in black (dashed line), $J_\perp(t)$ in green (solid line).}
\label{fig:Caso_2}
\end{figure}
The plots of $n_1(t)$ and $m_1(t)$, up to quantum fluctuations, show the periodic tunnelling of the weakly-populated vortex between the two rings, while the plots of $n_{-1}(t)$ and $m_{-1}(t)$ show the fluctuations of the vacuum state. Notice that chiral and rung currents' phases are permanently shifted of $\pi/2$, one being maximum (or minimum) when the other is zero. They somehow play a complementary role, analogous to that of position and momentum in a harmonic oscillator. This statement is exact in the limit $u\rightarrow 0$, a case where the expressions of chiral and rung currents simplify as follows: 
$$
    J_{chir}(t) = \sqrt{3}n_1(0) T_\parallel \cos \left(\frac{2tT}{\hbar} \right),
$$
$$
    J_\perp(t) = -n_1(0) T \sin \left(\frac{2tT}{\hbar} \right).
$$

Notice that, the bigger the value of parameter $T$, the wider the oscillations of the rung current, and the higher the frequencies of $J_{chir}$ and $J_\perp$. In short, a big value of $T$ is linked to a fast and efficient transfer of bosons between the two rings.

The presence of a non zero on-site repulsion is responsible for the periodic creation and annihilation of excited bosons pairs, which correspond to the high-frequency ripple in $n_{\pm 1}(t)$, $m_{\pm 1}(t)$ and $J_\perp(t)$ (see Figure \ref{fig:Caso_2}). As a consequence, the bigger the value of $u$, the wider the high-frequency oscillations of excited populations (and the smaller their period). This is a crucial point in order to obtain a both realistic and reliable description of the system: the global maximum of the excited populations must always be much less than the total number of bosons present in the system, otherwise the Bogoliubov approximation is invalidated and model's previsions turn unphysical. According to the analysis we have carried out, the Bogoliubov approximation ceases to be valid for relatively small values of $U/T_\parallel$ and surely \textit{before} approaching Mott's-lobes borders \cite{DosSantos}.

\begin{figure}[h!]
\includegraphics[width=1\columnwidth]{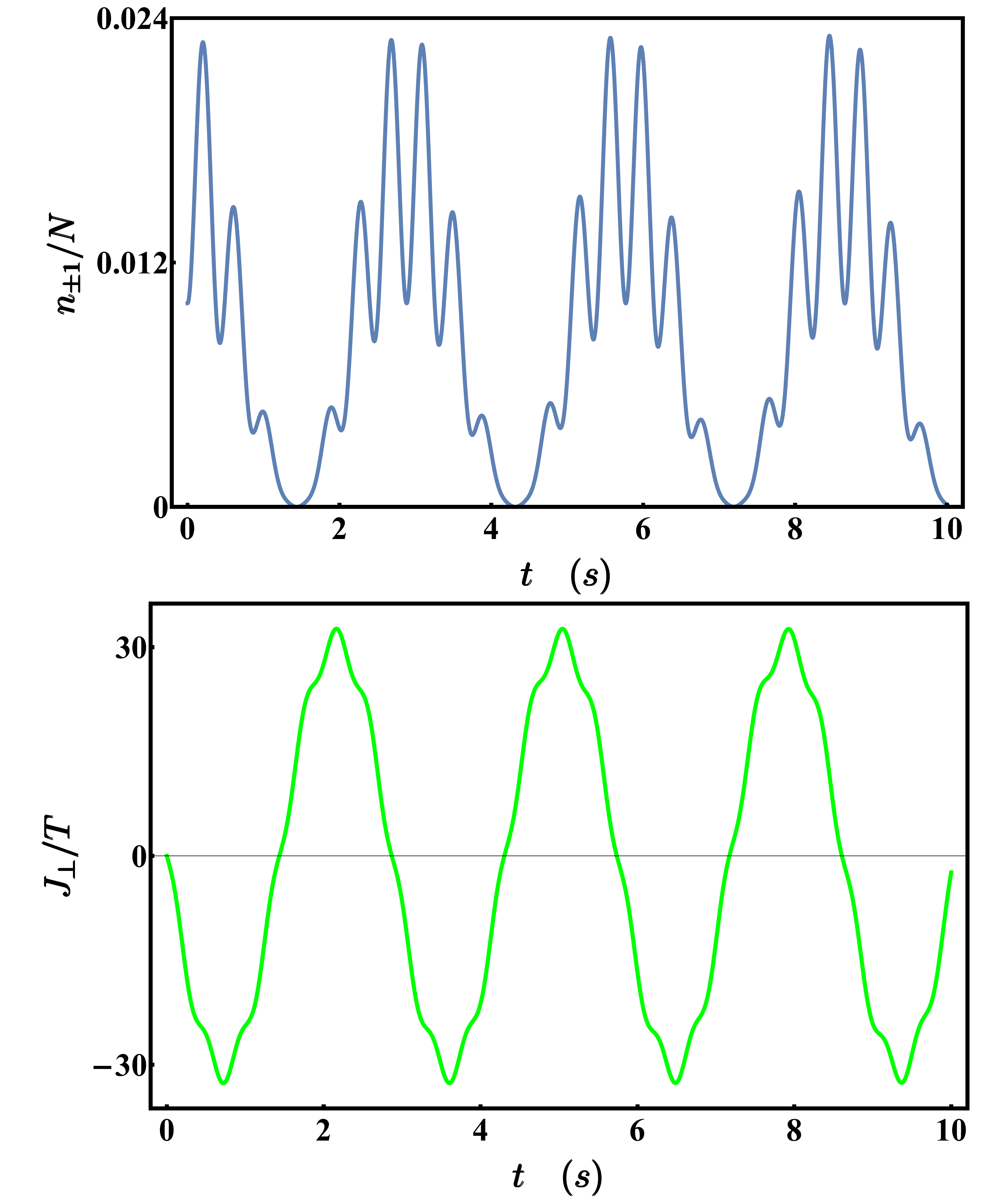}
\centering
\caption{Populations and current dynamics for $T_\parallel=2$, $U=0.01$, $T=1$, $\hbar=1$, $N=1000$, $n_1(0)=n_{-1}(0)=10$ and $m_{\pm 1}(0)=0$. Since the tunnelling process always involves \textit{pairs} of bosons, there cannot be momentum transfer between the two rings, hence chiral current is identically zero. $m_1(t)$ and $m_{-1}(t)$ feature the same behaviour of $n_{\pm 1}(t)$ but are shifted of a semi-period.}
\label{fig:Caso_3_1}
\end{figure}

\begin{figure}[t]
\includegraphics[width=1\columnwidth]{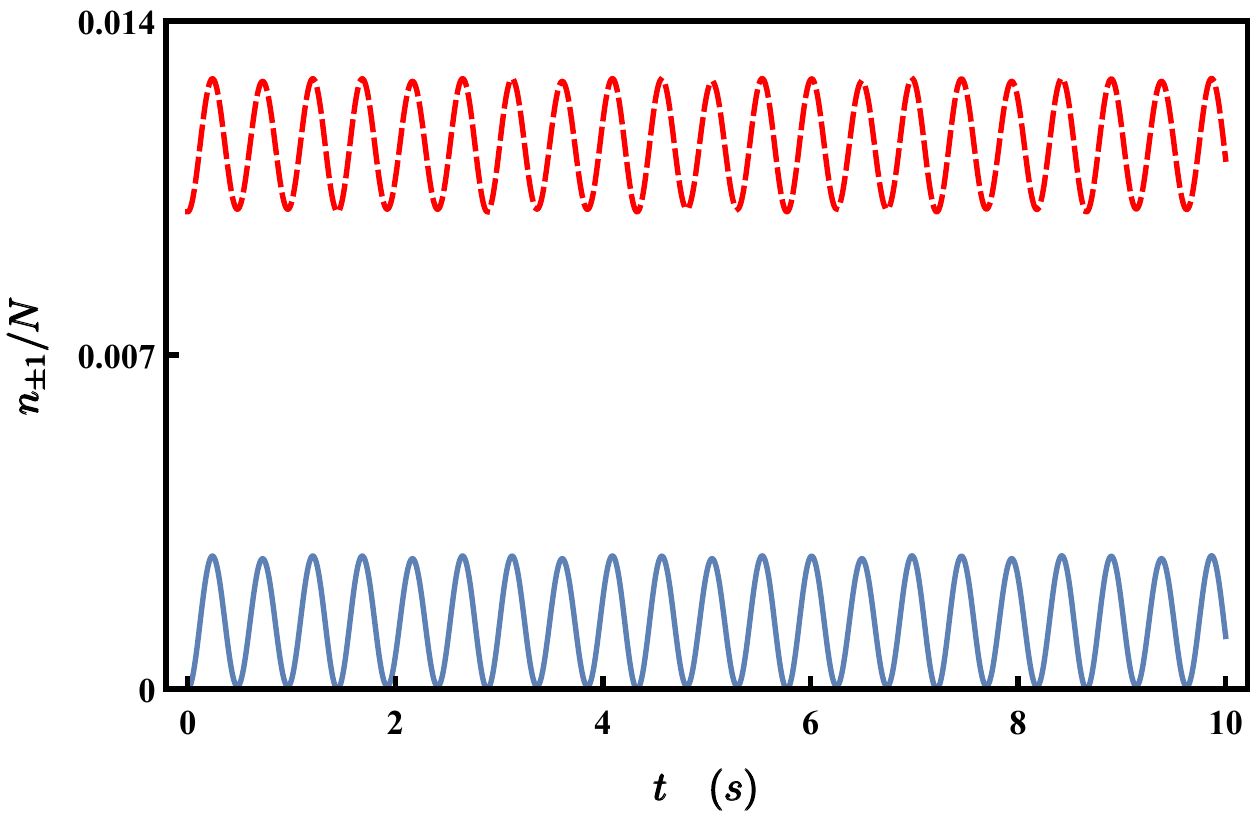}
\centering
\caption{Populations dynamics for $T_\parallel=2$, $U=0.01$, $T=1$, $\hbar=1$, $N=1000$, $n_1(0)=m_1(0)=10$, $n_{-1}(0)=m_{- 1}(0)=0$ and $\phi_1=\psi_1=0$. $n_{1}(t)$ corresponds to the red dashed line while $n_{-1}(t)$ corresponds to blue solid line. As the tunnelling is suppressed, $m_1(t)$ and $m_{-1}(t)$ feature the same behaviour (fluctuations) of $n_1(t)$ and $n_{-1}(t)$ respectively. The chiral and the rung current are identically zero.}
\label{fig:Caso_4_1}
\end{figure}

\begin{figure}[h!]
\includegraphics[width=1\columnwidth]{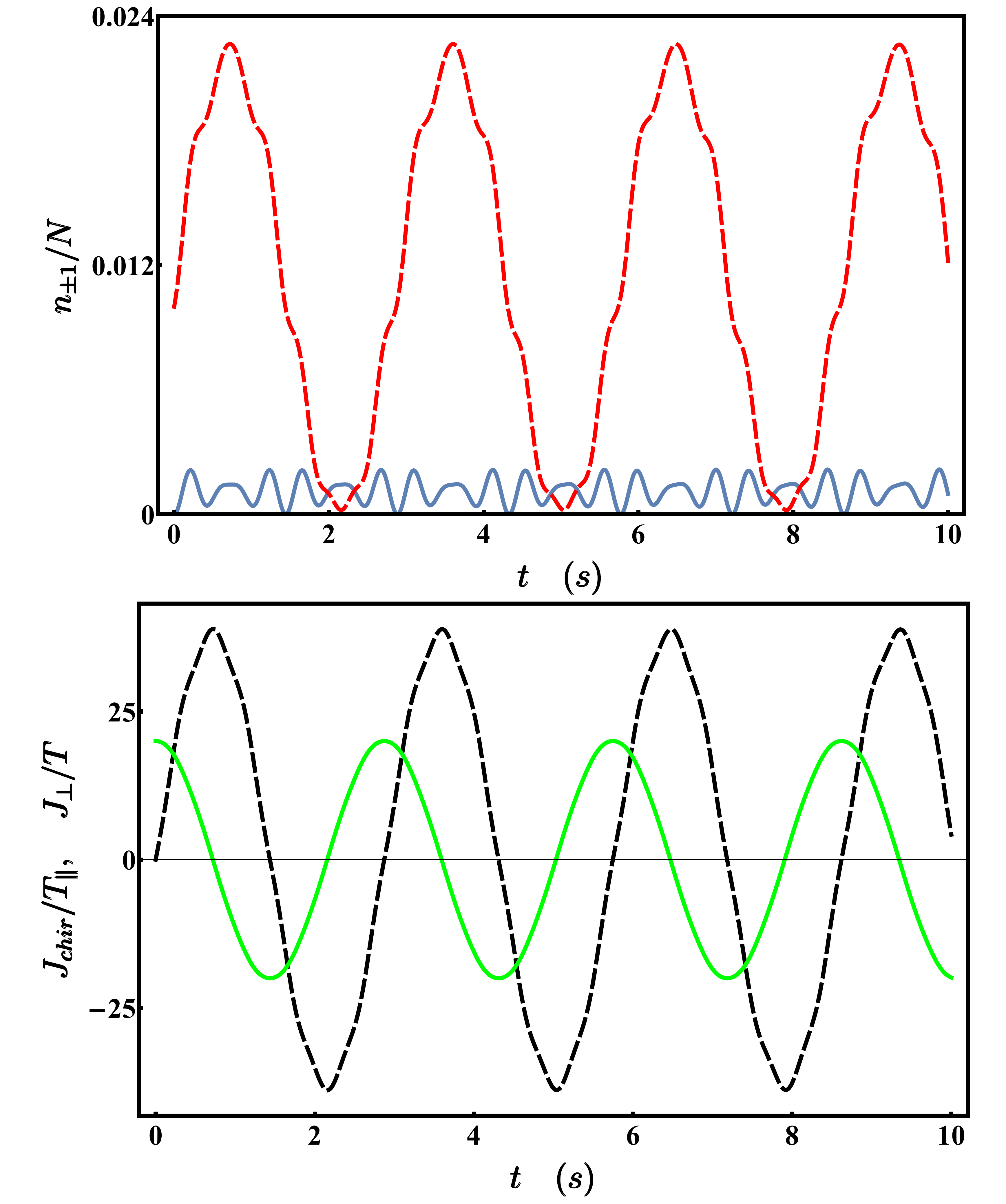}
\centering
\caption{Populations and currents dynamics for $T_\parallel=2$, $U=0.01$, $T=1$, $\hbar=1$, $N=1000$, $n_1(0)=m_1(0)=10$, $n_{-1}(0)=m_{- 1}(0)=0$ and $\psi_1=0$ but $\phi_1=\frac{\pi}{2}$. Upper panel: $n_{1}(t)$ corresponds to the red dashed line, while $n_{-1}(t)$ is depicted with the blue solid line. $m_{-1}(t)$ features exactly the same behaviour of $n_{-1}(t)$ while $m_1(t)$ is shifted of a semi-period with respect to $n_1(t)$. Apart from quantum fluctuations, one can notice that the weakly populated vortex with $k=+1$ periodically \textit{completely} transfers from one ring to the other and vice versa. Lower panel: Chiral current (in black), and rung current (in green).}
\label{fig:Caso_4_2}
\end{figure}

Moreover, it is interesting to notice that also the inter-ring tunnelling parameter $T$ affects the amplitude of excited-populations' oscillations and, when it approaches a certain upper limiting value, leads to dynamical instability. This issue will be deepened in next sub-section. 
\paragraph{Weak vortex and equal weak anti vortex in one ring, no excitations in the other ring.} If, at $t=0$, $m_1=m_{-1}=0$, but $n_1=n_{-1}\neq 0$ meaning that the initial state exhibits a balanced weak vortex-antivortex pair, as time goes on, excited bosons periodically tunnel from the first ring to the second ring and vice versa, giving place to a periodic rung current. Notice that the chiral current is identically zero, as the inter-ring tunnelling process always involves \textit{pairs} of bosons, as depicted in Figure \ref{fig:Caso_3_1}.
As regards the populations time-evolution, it is possible to recognize a low-frequency component, which corresponds to the periodic tunnelling of excited bosons between the two rings and a high-frequency component which corresponds to quantum fluctuations of the vacuum state, which in turn are are caused by a non-vanishing $u$. In this respect, notice that the on-site repulsion term $u$ is associated to two-bosons processes $a_1^\dagger a_{-1}^\dagger,\, a_1a_{-1},\, b_1^\dagger b_{-1}^\dagger $ and $b_1b_{-1}$, where the momentum in each single ring is indeed conserved. 
\paragraph{Two equal weak vortices in the two rings.} Let us assume that $n_{-1}(0)=m_{-1}(0)=0$ and that $n_1(0)=m_1(0)\neq 0$. This is a very interesting situation because the dynamics of our system inherently depends on the initial \textit{phase difference} $\phi_1(0)-\psi_1(0)$. If this difference is zero than the inter-ring tunnelling process is suppressed, and, up to quantum fluctuations, each ring always hosts the same number of excited bosons. As a consequence, chiral and rung current are identically zero (see Figure \ref{fig:Caso_4_1}). 
Conversely, a non-vanishing phase difference is responsible for a periodic transfer of excited bosons from one ring to the other and vice versa (See Figure \ref{fig:Caso_4_2}).
Hence, by observing $n_1(t)$ and $m_1(t)$ one can infer information about the \textit{phases} of the two weakly-populated vortices in the two rings. In this sense, the collective behavior which emerges in such a configuration can be used as a quantum interferometer.

\begin{figure}[h!]
\includegraphics[width=1\columnwidth]{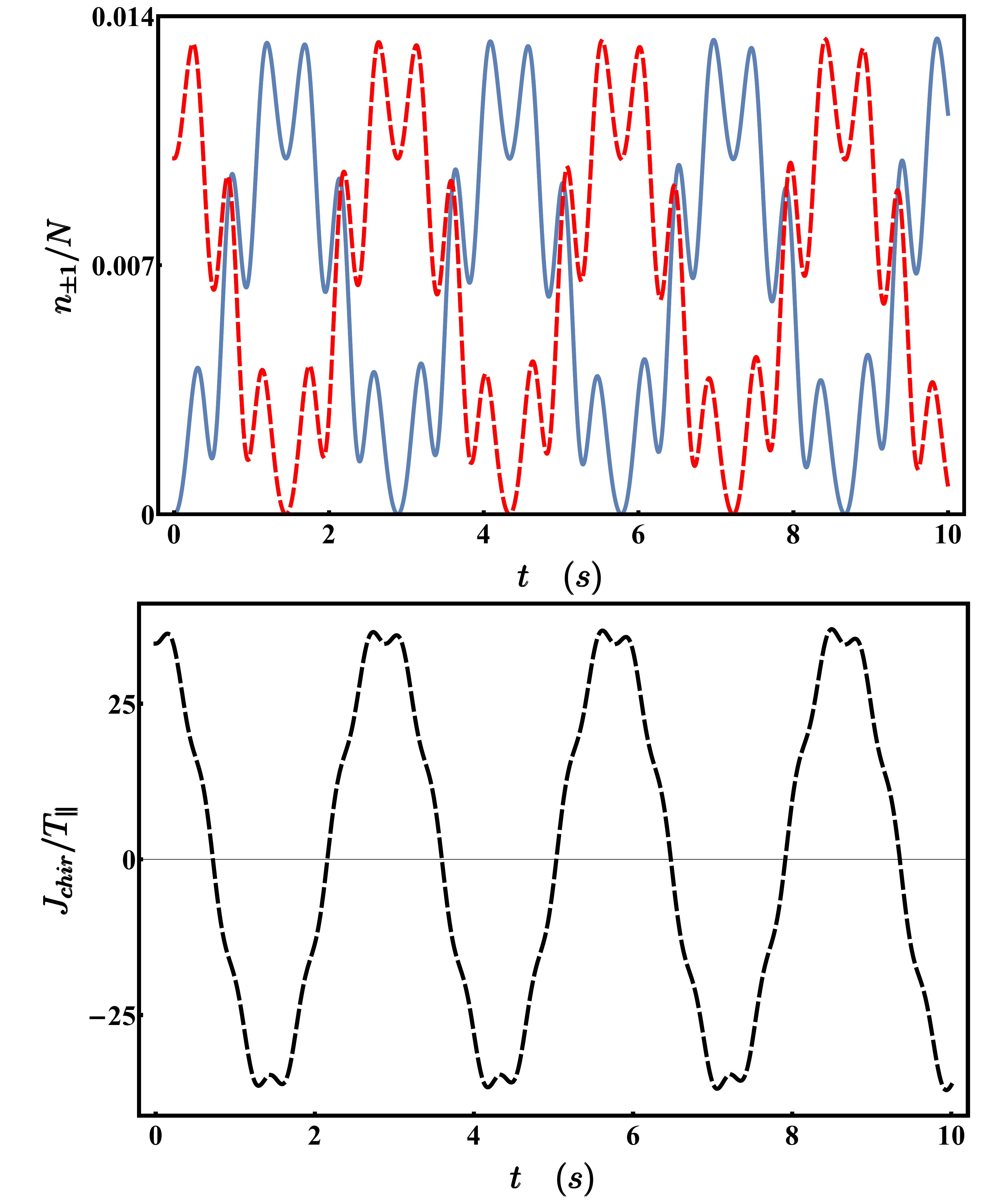}
\centering
\caption{Populations and current dynamics for $T_\parallel=2$, $U=0.01$, $T=1$, $\hbar=1$, $N=1000$, $n_1(0)=m_{-1}(0)=10$, $n_{-1}(0)=m_{ 1}(0)=0$. $n_1$ and $m_{-1}$ (dashed red line) have the same time evolution, and so $m_1$ and $n_{-1}$ (blue solid line). Rung current is identically zero.}
\label{fig:Caso_5_1}
\end{figure}

\paragraph{Weak vortex in one ring and equal (and opposite) weak antivortex in the other ring.} Let us assume that $n_{1}(0)=m_{-1}(0)\neq0$ and that $n_{-1}(0)=m_1(0)= 0$. Apart from quantum fluctuations (which correspond, as usual, to the high-frequency ripple in Figure \ref{fig:Caso_5_1}) excited bosons periodically tunnel from one ring to the other and vice versa. Remarkably, at each time, there are as many bosons which tunnel from ring A to ring B, as bosons which tunnel from ring B to ring A. As a consequence, there is a continuous momentum transfer between the two rings, i.e. a periodic $J_{chir}$ but, due to the symmetry of this tunnelling process, $J_\perp$ is identically zero (see Figure \ref{fig:Caso_5_1}). 

\subsection{Towards instability}
\label{sub:Instability}
\begin{figure}[h!]
\includegraphics[width=1\columnwidth]{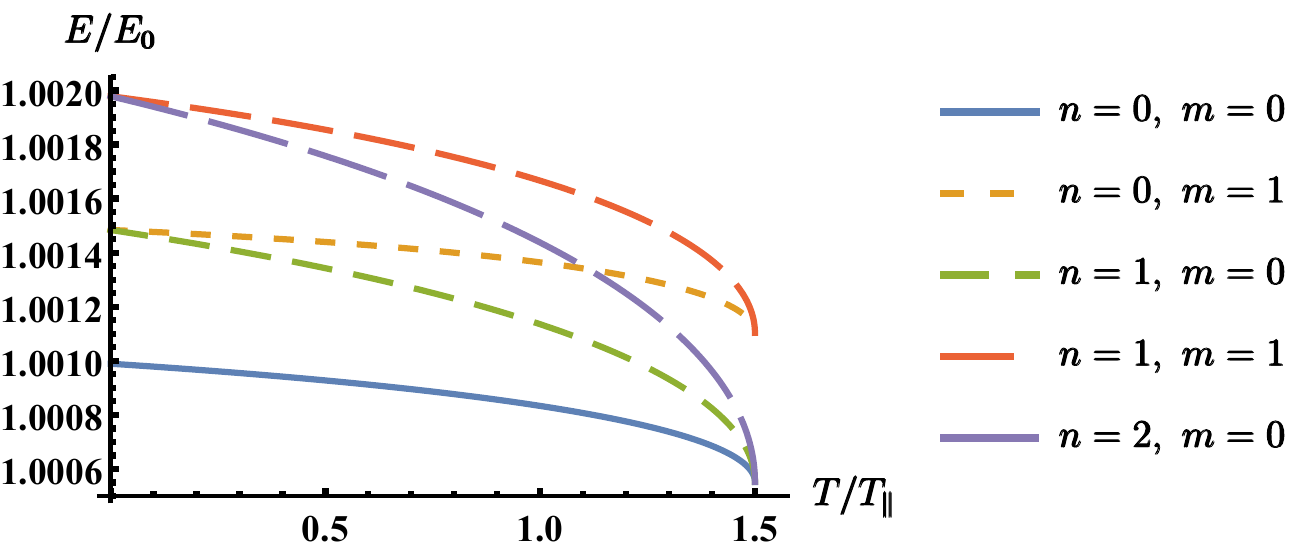}
\centering
\caption{Energy levels as a function of the inter tunnelling parameter $T$. When $T=0$ the rings are decoupled, when $T\rightarrow \frac{3}{2}T_\parallel$ there is the spectral collapse with respect to the first generalized harmonic oscillator. }
\label{fig:Spectral_collapse}
\end{figure}

The diagonalization scheme presented in Section \ref{sec:Double_trimer} shows that there are some \textit{constraints} on generalized rotation angles $\theta_a$ and $\theta_b$ (see Equations (\ref{eq:Rotation_angles})). The same constrains also recur in the expression of diagonal Hamiltonian (\ref{eq:Diagonal_Hamiltonian}). Recalling that $T_\parallel$, $u$ and $T$ are, by definition, non-negative numbers, all the diagonalization scheme and the dynamical results that we have presented so far are well defined iff $T<\frac{3}{2}T_\parallel$. If the inter-ring tunnelling parameter $T$ becomes so large to approach the limiting value $\frac{3}{2}T_\parallel$, one can observe the spectral collapse, meaning that the separation between subsequent energy levels tends to zero (see Figure \ref{fig:Spectral_collapse}).

In this respect, one should recall that diagonal Hamiltonian $\hat{\mathcal{H}}$ is, up to a constant term $E_0$, the sum of \textit{four} harmonic oscillators, two of them having frequency $\Omega$, and the others $\omega$ (see equations (\ref{eq:A3B3}) and (\ref{eq:Diagonal_Hamiltonian})).
As a consequence, two integer quantum numbers ${n}=n_1+n_{-1}$ and ${m}=m_1+m_{-1}$ are enough to label the energy levels of the system
$$
\hat{\mathcal{H}}=E_0 +\hbar\Omega\, \left({n} +1 \right)+ \hbar \omega\,\left( {m}+1\right).
$$
Of course an energy level labelled by quantum numbers $\left({n},{m}\right)$ is $({n}+1)({m}+1)$ times degenerate. This is the number of eigenstates
$|n_1,n_{-1},m_1,m_{-1}\rangle$ associated to a given energy.  
Figure \ref{fig:Spectral_collapse} clearly shows that, if $T\rightarrow 0$, then $\Omega \rightarrow \omega$, meaning that one has the spectrum of a harmonic oscillator of frequency $\omega$ whose levels are $({m}+3)!/({m}!\,3!)$ times degenerate. The same figure well illustrates the spectral collapse of (the energy levels of) the $\Omega$-dependent harmonic oscillator for $T\rightarrow \frac{3}{2}T_\parallel$. 

\begin{figure}[h!]
\includegraphics[width=1\columnwidth]{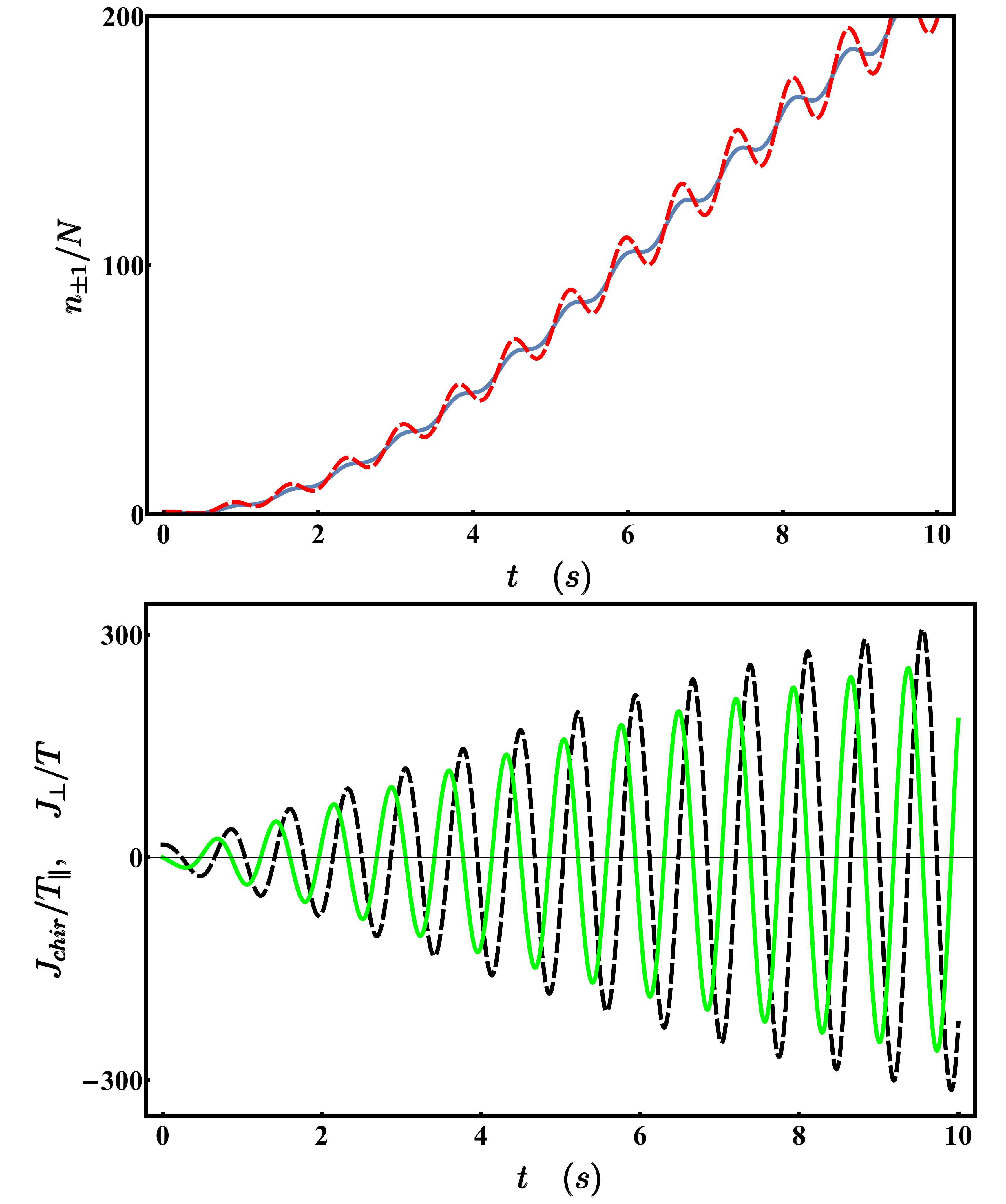}
\centering
\caption{Populations and currents dynamics for $T_\parallel=2$, $U=0.01$, $T=2.999$, $\hbar=1$, $N=1000$, $n_1(0)=10$ and $n_{-1}(0)=m_{\pm 1}(0)=0$. Upper panel: $n_{1}(t)$ is corresponds to the red dashed line and $n_{-1}(t)$ is depicted with a blue solid line. $m_{1}(t)$ and $m_{-1}(t)$ essentially feature the same diverging behaviour. The number of excited bosons rapidly increases and soon becomes nonphysical. Lower panel: chiral (black dashed line) and rung (green solid line) currents exhibit a divergence too.}
\label{fig:Divergence}
\end{figure}

As regards excited populations, approaching the border of the \textit{stability} region, one observes 
that the numbers of excited bosons feature a \textit{diverging} behavior, as Figure \ref{fig:Divergence} clearly depicts. 
This circumstance is not just a mathematical accident, but serves to set the validity 
range of our model. Moreover, since the spectral collapse of the energy levels and the
divergence of physical observables typically marks the appearance of unstable regimes, this 
phenomenology suggests the presence of a dynamical phase transition \cite{Penna_central_depleted}. 
A well-known example is supplied by the BH ring model with attractive bosons
where the interplay of the hopping parameter with the negative interaction can
cause the spectral collapse \cite{Jack}.
In this respect, we emphasize how the presence of the inter-ring boson exchange described by $T$
allows one to trigger unstable behaviors in the current model. 
The case $T>{3}T_\parallel / {2}$ will be explored in a separate paper.


\subsection{Instability in a general double ring} 
The limiting condition $T< {3}T_\parallel / {2}$ has been derived with reference to the double trimer. 
Nevertheless, it is possible to derive an analogous stability condition for a double ring which features 
a general number of sites $M_s$. Considering formula (\ref{eq:decoupling}), one notices that each 
sub-Hamiltonian $\hat{H}_k$ features two characteristic frequencies
\begin{equation}
\omega_k = \frac{1}{\hbar}\sqrt{\left(2T_\parallel C_k+2u\right)2T_\parallel C_k },
\label{omegak}
\end{equation}
\begin{equation}
\Omega_k = \frac{1}{\hbar}\sqrt{\left(2T_\parallel C_k-2T\right)\left(2T_\parallel C_k +2u-2T\right)},
\label{Omegak}
\end{equation}
where $C_k:=1- \cos \left( {2\pi} k/{M_s} \right)$. In passing, can observe that, choosing $M_s=3$ and 
$k=1$, one re-obtains characteristic frequencies (\ref{eq:Char_freq_1}) and (\ref{eq:Char_freq_2}). 
Recalling that $T_\parallel$, $u$ and $T$ are \textit{positive} parameters, the stability condition 
is given by
$$
  2T_\parallel C_k -2T > 0 ,
$$
which results in
$$
 T<T_\parallel C_k .
$$
Since $C_k$ is a monotonic increasing function for 
$0<k<{(M_s-1)}/{2}$, the limiting value of $T$ is found for $k=1$, i.e.
$$
 T<T_\parallel\left[1-\cos \left(\frac{2\pi}{M_s}\right)\right].
$$
In other words, whatever the number of sites in the system, the sub-Hamiltonian which collapses first (due to an increase of $T$) is always $\hat{H}_1$. 

As already mentioned, our model can describe the dynamics also of systems which feature an \textit{attractive} interaction $U<0$, provided that condition $|U|/T$ small enough is fulfilled, thus guaranteeing that bosons are superfluid and delocalized \cite{Jack}. For such systems, two conditions
$$
    \left\lbrace
\begin{array}{ll}
 u > T_\parallel \left[ \cos \left( \frac{2\pi}{M_s} \right)-1 \right] \\
 \\
 T<T_\parallel\left[1-\cos \left(\frac{2\pi}{M_s}\right)\right] +u 
\end{array}
\right.
$$
ensure that the spectrum is real and discrete.
A systematic exploration of the attractive regime includes the case when $|U |/T$ is large.
In this case the formation of soliton-like quantum states characterized by boson localization
(see, for example \cite{Vezzani}, \cite{Kavoulakis} and \cite{Kanamoto}) typically occurring in
the single-ring geometry, is expected. The quantum study of the interaction between solitons
on different rings will be developed elsewhere along the lines of reference \cite{Cavaletto} 

\section{Concluding remarks}
\label{sec:Concluding_remarks}
In this work we have focused on the BH two-ring ladder. In Section \ref{sec:Model_presentation} 
we have shown that, moving to momentum-modes picture and performing the well known Bogoliubov approximation, 
the system Hamiltonian, up to a constant term $E_0$, decouples in $(M_s-1)/2$ independent Hamiltonians 
$\hat{H}_k$, one for each \textit{pair} of momentum modes. In Section \ref{sec:Dynamical_algebra}, we have 
proved that each Hamiltonian $\hat{H}_k$ belongs to a dynamical algebra so(2,3). This property has provided 
not only an effective diagonalization scheme, but also the possibility of computing the
conserved quantity in the system and its dynamical equations. 

Section \ref{sec:Double_trimer} has been devoted to apply our solution scheme to a simple and yet very 
interesting example: the double twin 
trimer where the ground state features a $r=0$ mode macroscopically occupied. After finding the explicit 
expression of its spectrum, we have shown that the excitations of the system indeed can be seen as 
weakly-populated vortices. Then we have introduced some significant physical observables, which are
currently used in literature (\cite{Natu}, \cite{Piraud}) and computed their time-evolution thanks to the 
closure property of the algebraic framework. 

The derived dynamical equations have highlighted 
the fundamental processes which happen in the system. We have also noticed that, while some 
``global" observables (namely $N_*$ and $J_\perp$) can be written as elements of the dynamical
algebra so(2,3), in order to have a more detailed description of microscopic physical processes
(e.g., the time evolution of boson populations) 
it is necessary to perform the immersion of algebra so(2,3) in the larger  
15-dimensional algebra so(2,4). It is within this larger algebraic framework that the dynamics of 
all the observables typically used in literature takes place. 

Finally, in Section \ref{sec:Vortex_dynamics}, 
we have explored the system evolution for different choice of parameters and initial conditions. 
In particular, we have explicitly described the vacuum-state fluctuations and the coherent time-evolution
of the rung and the chiral currents. Also,  we have found a configuration where a different choice in 
the initial \textit{phase difference} of the excitation modes in the two rings 
allows to completely inhibit the boson inter-ring exchange. As a conclusion, 
we have analysed the stability region of the system, and noticed that, for $T\to \frac{3}{2}T_\parallel$,
the system turns to be unstable, hinting the possible presence of a dynamical phase transition. This 
issue, the study of strongly interacting attractive bosons, and the study of the excitation dynamics for the macroscopic modes $r \ne 0$ (vortex configurations)
will be explored in a future work.


\begin{appendices}

\section{}
\label{sec:Bogoliubov_approximation}
The two terms with the triple summation can be simplified considering just the addends which include at least two $r$-mode operators:   
$$
\sum_{p,q,k=1}^{M_s}    a_{q+k}^\dagger a_{p-k}^\dagger a_q a_{p} \approx n_r(n_r-1) + 4n_r \sum_{k\neq r}n_k
$$
$$
 + \left(a_r\right)^2 \sum_{k\neq r} a^\dagger_{r+k} a^\dagger_{r-k} +  \left(a_r^\dagger \right)^2 \sum_{k\neq r} a_{r+k} a_{r-k}
$$
Of course, the same reasoning holds also for the second ring, i.e. for the term proportional to $U_b$.
\\
\noindent 
According to the well known Bogoliubov approximation, a mode operator relevant to a macroscopically occupied mode can be declassed to a complex number whose phase can be arbitrarily chosen to be zero. Performing the substitutions $a_r^\dagger \rightarrow  \sqrt{n_r}$, $a_r  \rightarrow \sqrt{n_r}$, $b_r^\dagger  \rightarrow  \sqrt{m_r}$, $b_r  \rightarrow  \sqrt{m_r}$ and writing $n_r$ as $N-\sum_{k\neq r} n_k$ and $m_r$ as $M-\sum_{k\neq r} m_k$, the Hamiltonian assumes the following form: 
$$
\hat{H}=\hat{H}_a + \hat{H}_b + \hat{H}_\perp
$$
where
$$
    \hat{H}_a = -2T_a \biggl[N\cos(a\tilde{r}) + \sum_{k\neq r}\left(\cos(a\tilde{k})-\cos(a\tilde{r})\right)n_k  \biggr]+
$$
$$
 \frac{u_a }{2} \biggl[ N -1 + \sum_{k\neq 0} \left(n_{r+k} +n_{r-k} +a_{r+k}^\dagger a_{r-k}^\dagger+ a_{r+k} a_{r-k} \right)\biggr]  
$$
$$
    \hat{H}_b = -2T_b \biggl[M\cos(a\tilde{r}) + \sum_{k\neq r}\left(\cos(a\tilde{k})-\cos(a\tilde{r})\right)m_k  \biggr]+
$$
$$
 \frac{u_b}{2} \biggl[ M -1 + \sum_{k\neq 0} \left(m_{r+k} +m_{r-k} +b_{r+k}^\dagger b_{r-k}^\dagger+ b_{r+k} b_{r-k} \right)\biggr]  
$$
$$
\hat{H}_\perp  = -2T\sqrt{NM} -T\sqrt{\frac{M}{N}} \sum_{k\neq r} n_k -T \sqrt{\frac{N}{M}} \sum_{k\neq r} m_k+
$$
$$
\frac{T}{2}\sum_{k\neq r} \left(a_{r+k}b_{r+k}^\dagger +a_{r+k}^\dagger b_{r+k} + a_{r-k}b_{r-k}^\dagger +a_{r-k}^\dagger b_{r-k} \right)
$$

\section{}
\label{sec:Defining_commutators}
Here we give the defining commutators of our 10-dimensional algebra. In the text, we have already mentioned that $\left\{A_+,\,A_-,\,A_3\right\}$ generate an su(1,1) algebra, and $\left\{B_+,\,B_-,\,B_3\right\}$ too. Concerning operators $\left\{S_+,\,S_-,\,A_3-B_3\right\}$, they generate an su(2) algebra, the latter being marked by commutators: $\left[S_+,\,S_-\right]=2(A_3-B_3)$, $\left[S_\pm,\, A_3-B_3 \right]=\mp S_\pm$. Operators $\left\{K_+,\,K_-,\,A_3+B_3\right\}$ generate another su(1,1) algebra, the defining commutators being: $\left[K_+,\,K_- \right]=-2\left(A_3+B_3\right)$, $\left[K_\pm, \, A_3+B_3\right]=\mp K_\pm$. After recognizing the presence of these $4$ sub-algebras, we give the commutation relations between the elements of the various sub-algebras. Any element $A_i$ commutes with all elements $B_j$, i.e. $[A_i,B_j]=0$, with $i,j=+,-,3$. Moreover
$$
\left[A_+ ,\,S_+ \right]=0   ,\quad
\left[A_+ ,\,S_- \right]=-K_+   ,\quad 
\left[A_+ ,\,K_+ \right]=0,   
$$
$$
\left[A_+ ,\,K_- \right]=-S_+,   \quad
\left[A_- ,\,S_+ \right]=K_-,   \quad 
\left[A_- ,\,S_- \right]=0,  
$$
$$
\left[A_- ,\,K_+ \right]=S_-,    \qquad
\left[A_- ,\,K_- \right]=0,   
$$
$$
\left[A_3,S_\pm\right]=\pm\frac{1}{2}S_\pm, \qquad 
\left[A_3,K_\pm\right]=\pm\frac{1}{2}K_\pm,
$$

$$
\left[B_+ ,\,S_+ \right]=-K_+   ,\quad
\left[B_+ ,\,S_- \right]=0   ,\quad 
\left[B_+ ,\,K_+ \right]=0,   
$$
$$
\left[B_+ ,\,K_- \right]=-S_-,   \quad
\left[B_- ,\,S_+ \right]=0,   \quad 
\left[B_- ,\,S_- \right]=K_-,  
$$
$$
\left[B_- ,\,K_+ \right]=S_+,    \qquad
\left[B_- ,\,K_- \right]=0,   
$$
$$
\left[B_3,S_\pm\right]=\mp\frac{1}{2}S_\pm, \qquad 
\left[B_3,K_\pm\right]=\pm\frac{1}{2}K_\pm, \qquad
$$
$$
\left[S_+  ,\, K_+   \right]= 2A_+, \qquad 
\left[S_+   ,\,K_-   \right]=-2B_-, 
$$
$$
\left[S_-  ,\, K_+   \right]= 2B_+, \qquad 
\left[S_-   ,\,K_-   \right]=-2A_- 
$$

\onecolumngrid
\section{}
\label{sec:Rotation_angles}
The explicit expressions of the unitary transformations have been computed making use of the well known Campbell-Baker-Hausdorf formula 
$$
  e^X\,Y\,e^{-X}= \sum_{k=0}^{+\infty}\frac{1}{k!}[X,Y]_k
$$
As we are working within an algebraic framework, algebra's closure property guarantees that any commutator of two algebra elements is still an algebra element. Generalized rotation angles $\varphi$, $\xi$, $\theta_a$ and $\theta_b$ are computed by imposing the nullification of non-diagonal operators. Their expression is

\bigskip

$$
{\rm tan} \varphi = 
\frac{2T(\gamma_{a,k}+\gamma_{b,k})}{(u_a+u_b)(u_a-u_b)-(\gamma_{a,k}+\gamma_{b,k})(\gamma_{a,k}-\gamma_{b,k}) }
$$
$$
{\rm th} \xi = \frac{2T(u_b-u_a)}{(u_a^2-u_b^2-\gamma_{a,k}^2 +\gamma_{b,k}^2)\sqrt{1+ \chi }} ,
\qquad
\chi = \frac{ 4T^2( \gamma_{a,k}+\gamma_{b,k})^2 }{ (u_a^2-u_b^2-\gamma_{a,k}^2 +\gamma_{b,k}^2)^2 }
$$
$$
{\rm th} \theta_a = \frac{c_2}{c_1}, \qquad  \qquad {\rm th} \theta_b = \frac{c_4}{c_3}
$$
Coefficients $c_1$, $c_2$, $c_3$ and $c_4$ have the following expressions

$$
 c_1=(u_a-u_b) \sinh \xi \sin \varphi + \left(\gamma_{a,k}+ \gamma_{b,k} \right) \cosh \xi -2 T \sin \varphi+ (\gamma_{a,k}-\gamma_{b,k}) \cos \varphi 
$$
$$
    c_2= 2T \sinh \xi \cos \varphi+(\gamma_{a,k}-\gamma_{b,k})\sinh \xi \sin \varphi + \cosh \xi(u_a+u_b)+ (u_a-u_b)\cos \varphi 
$$
$$
 c_3=(u_a-u_b) \sinh \xi \sin \varphi + \left(\gamma_{a,k}+ \gamma_{b,k} \right) \cosh \xi +2 T \sin \varphi- (\gamma_{a,k}-\gamma_{b,k}) \cos \varphi 
$$
$$
    c_4= 2T \sinh \xi \cos \varphi+(\gamma_{a,k}-\gamma_{b,k})\sinh \xi \sin \varphi +  \cosh \xi(u_a+u_b)- (u_a-u_b)\cos \varphi 
$$

\section{}
\label{sec:Commutators_so(2,4)}
The fifteen generators of algebra so(2,4) are
$$
 A_+ = a_1^\dagger a_{-1}^\dagger ,
 \quad
 A_- = A_+^\dagger,
 \quad
 A_3 = \frac{1}{2}(n_1+n_{-1}+1),
 \qquad
 B_+ =  b_1^\dagger b_{-1}^\dagger,
 \quad
 B_- = B_+^\dagger,
 \quad
 B_3 = \frac{1}{2} (m_1+m_{-1}+1),
$$
$$
 Q_+ = a_1^\dagger b_1,
 \quad
 Q_- = Q_+^\dagger 
 \qquad
 G_+ = a_{-1}^\dagger b_1^\dagger,
 \quad
 G_- = G_+^\dagger
$$
$$
 H_+ = a_1^\dagger b_{-1}^\dagger  ,
 \quad
 H_- = H_+^\dagger
 \qquad
 R_+ = a_{-1}^\dagger b_{-1},
 \quad
 R_- = R_+^\dagger , 
 $$
 $$
 T = \frac{1}{2} \left [ (n_1-n_{-1} )- ( m_1-m_{-1} ) \right ]
$$
The relevant Casimir operator is 
$$
C= A_3^2 +B_3^2 +\frac{1}{2} \; T^2 
%
%
+\frac{1}{2}\biggl[ Q_+Q_-+ Q_- Q_+ + R_+R_-+ R_- R_+ 
\biggr]
$$
$$
-\frac{1}{2}\biggl[ A_+A_-+ A_- A_+ + B_+ B_-+ B_- B_+ G_+G_-+ G_- G_+ + H_+ H_-+ H_- H_+
\biggr],
$$
which can be written in the standard form $C =\frac{3}{2} K_4\left(K_4+2\right)$ where 
$K_4= ({n_1-n_{-1}+m_1-m_{-1}-2})/{2}$ represents the total angular momentum. This shows how the algebra so(2,4)
again features the total angular momentum as a conserved quantity.
\end{appendices}

\vfill
\eject

\twocolumngrid

\end{document}